\documentclass[11pt,a4paper]{article}
\usepackage{jcappub,natbib}
\usepackage{braket}
\input{colordvi.tex}
\title{Polarization bispectrum for measuring primordial magnetic fields}
\author[a,b]{Maresuke Shiraishi}
\affiliation[a]{Dipartimento di Fisica e Astronomia ``G. Galilei'', Universit\`a degli Studi di Padova, via Marzolo 8, I-35131, Padova, Italy}
\affiliation[b]{INFN, Sezione di Padova, via Marzolo 8, I-35131, Padova, Italy}
\emailAdd{maresuke.shiraishi@pd.infn.it}

\abstract{%
We examine the potential of polarization bispectra of the cosmic microwave background (CMB) to constrain primordial magnetic fields (PMFs). We compute all possible bispectra between temperature and polarization anisotropies sourced by PMFs and show that they are weakly correlated with well-known local-type and secondary ISW-lensing bispectra. From a Fisher analysis it is found that, owing to E-mode bispectra, in a cosmic-variance-limited experiment the expected uncertainty in the amplitude of magnetized bispectra is 80\% improved in comparison with an analysis in terms of temperature auto-bispectrum alone. In the {\it Planck} or the proposed PRISM experiment cases, we will be able to measure PMFs with strength 2.6 or 2.2 nG. PMFs also generate bispectra involving B-mode polarization, due to tensor-mode dependence. We also find that the B-mode bispectrum can reduce the uncertainty more drastically and hence PMFs comparable to or less than 1 nG may be measured in a PRISM-like experiment.
}

\begin{document}

\maketitle
\flushbottom

%%%%%%%%%%%%%%%%%%%%%%%%%%%%%%%%%%%%%%%
\section{Introduction}

Several cosmological and astrophysical observations support existence of finite magnetic fields in galaxies, cluster of galaxies or large voids (e.g., \cite{Bernet:2008qp, Wolfe:2008nk, Neronov:1900zz, Tavecchio:2010mk, Dolag:2010ni, Takahashi:2013uoa}). There are a variety of studies where these origin is linked with primordial vector field in the very early Universe (e.g., refs.~\cite{Bamba:2006ga, Martin:2007ue, Kanno:2009ei, Demozzi:2009fu}).\footnote{At the same time, several papers have also discussed possibilities of magnetic field production in the late-time Universe (e.g., refs.~\cite{Ichiki:2006cd, Maeda:2008dv, Fenu:2010kh, Hollenstein:2012mb, Saga:2013glg}).} Despite a fact that such models are strongly constrained by conditions not to contradict inflation or the high energy physics \cite{Suyama:2012wh, Demozzi:2012wh, Fujita:2012rb, Ringeval:2013hfa, Ferreira:2013sqa}, these provide phenomenologically interesting outputs (e.g., refs.~\cite{Seery:2008ms, Caldwell:2011ra, Urban:2012ib, Barnaby:2012tk, Motta:2012rn, Jain:2012ga, Shiraishi:2012xt, Jain:2012vm, Bartolo:2012sd, Shiraishi:2013sv, Shiraishi:2013vja, Kunze:2013hy}). 

In this paper, we focus on magnetized non-Gaussian signals in the cosmic microwave background (CMB). Primordial magnetic fields (PMFs) create not only scalar-mode but also vector-mode and tensor-mode CMB anisotropies via energy density and anisotropic stress fluctuations \cite{Subramanian:1998fn, Mack:2001gc, Subramanian:2002nh, Subramanian:2003sh, Giovannini:2005jw, Paoletti:2008ck, Yamazaki:2008gr,Shaw:2009nf, Giovannini:2009ts}. Recent analyses using CMB power spectra suggest nearly scale-invariant PMFs with strength less than about 3 nG \cite{Paoletti:2010rx, Shaw:2010ea, Paoletti:2012bb, Ade:2013lta}. On the other hand, under an assumption of Gaussianity of PMFs, CMB polyspectra also be generated due to quadratic dependence of the stress fluctuations on Gaussian PMFs \cite{Brown:2005kr, Seshadri:2009sy, Caprini:2009vk,Cai:2010uw, Shiraishi:2010yk, Kahniashvili:2010us, Shiraishi:2011dh, Shiraishi:2011fi, Shiraishi:2012rm, Trivedi:2010gi}. They have diverse shapes unlike CMB bispectra from standard scalar non-magnetized non-Gaussianities since in PMF case the vector-mode and tensor-mode non-Gaussianities can be enhanced \cite{Shiraishi:2010yk, Shiraishi:2011dh, Shiraishi:2011fi, Shiraishi:2012rm}. The magnetized bispectrum has provided a new observational constraint on PMFs consistent with bounds from the power spectrum \cite{Shiraishi:2013wua}. 

These previous studies have analyzed effects of temperature auto-bispectrum alone. On the other hand, it is known that polarization bispectra can also help to determine non-Gaussianity parameters \cite{Babich:2004yc, Yadav:2007rk, Yadav:2007ny} and they will be utilized in data analysis of the {\it Planck} or the proposed PRISM experiment \cite{:2006uk, Andre:2013afa}. In this sense, studying impacts of PMFs on the polarization bispectra will be useful and timely. 

On the basis of these motivations, this paper investigates the potential of the polarization bispectra sourced by PMFs. We compute magnetized auto- and cross-bispectra between temperature, E-mode and B-mode anisotropies, and forecast the uncertainty of the amplitude of these bispectra, which depends on PMF strength, via the Fisher analysis. As observations, we assume the {\it Planck} and the proposed PRISM experiments. In computation of the CMB bispectra, we consider the dependence on the scalar and tensor modes and ignore the vector mode because of its smallness on scales where we focus on. Then, we confirm that owing to tensor-mode contribution, the polarization bispectra reduce the uncertainty of the magnetized bispectra more drastically in comparison with a forecast from the temperature auto-bispectrum alone. We also find that the existence of local-type non-Gaussianity and secondary ISW-lensing signal does not bias an error estimation of the amplitude of the magnetized bispectra. We follow the formulae and computational procedure in ref.~\cite{Shiraishi:2012rm}.

This paper is organized as follows. In the next section, we analyze signatures of all possible magnetized bispectra composed of the temperature, E-mode and B-mode anisotropies. In section~\ref{sec:Fisher}, through the Fisher analysis, we discuss the detectability of the magnetized bispectra. The final section is devoted to summary and discussion of this paper. In appendices~\ref{appen:noise} and \ref{appen:fnl}, we summarize instrumental noise information utilized in section~\ref{sec:Fisher} and the uncertainty of the local-type non-Gaussianity. 

%%%%%%%%%%%%%%%%%%%%%%%%%%%%%%%%%%%%%%%%%%
\section{Temperature and polarization bispectra originating from primordial magnetic fields}

In this section, we examine the dependence of all possible temperature and polarization bispectra generated from PMFs for $\ell < 2000$. The notations and conventions are consistent with refs.~\cite{Shiraishi:2010yk, Shiraishi:2011fi, Shiraishi:2011dh, Shiraishi:2012rm}. 

%/////////////////////////////////////////
\subsection{Magnetized CMB fluctuation}

Let us start from a cosmological model with large-scale magnetic fields which are created at very early stages of the Universe and stretched beyond horizon by the inflationary expansion. With assumptions of Gaussianity of PMFs and their evolution like radiations: $B_i \propto a^{-2}$, the PMF power spectrum normalized at the present epoch is given as 
\begin{eqnarray}
\Braket{ B_i({\bf k}) B_j({\bf k'}) } 
= (2\pi)^3 \frac{P_B(k)}{2} P_{ij}(\hat{\bf k})\delta({\bf k} + {\bf k'})~,
\end{eqnarray}
where $P_{ij}(\hat{\bf k}) = \delta_{ij} - \hat{k}_i \hat{k}_j$ is a projection tensor which reflects the divergenceless nature of PMFs. The shape of $P_B(k)$ depends strongly on models of primordial magnetogenesis. In order to find observational clues, it is often parametrized as the power-law type:
\begin{eqnarray}
P_B(k) = A_B k^{n_B}~,
\end{eqnarray}
where the amplitude depends quadratically on the PMF strength smoothed on $r$ as 
\begin{eqnarray}
{A}_B = {
\left(2 \pi \right)^{n_B + 5} B_r^2 \over \Gamma(\frac{n_B + 3}{2}) k_r^{n_B + 3} }~.
\end{eqnarray}

PMFs create energy-momentum tensor as
\begin{eqnarray}
\begin{split}
T^i_{~j}({\bf k},\tau) &\equiv {\rho_\gamma}(\tau)
\left[ \delta^i_{~j} \Delta_B({\bf k}) + \Pi_{Bj}^i({\bf k}) \right]~, \\
\Delta_B({\bf k}) &= {1 \over 8\pi \rho_{\gamma,0}}
\int \frac{d^3 {\bf k}'}{(2\pi)^3} 
B^i({\bf k'}) B_j({\bf k} - {\bf k'}) ~, \\
\Pi^i_{Bj}({\bf k}) &=-{1 \over 4\pi \rho_{\gamma,0}} \int \frac{d^3
 {\bf k'}}{(2 \pi)^3} B^i({\bf k'}) B_j({\bf k} - {\bf k'})~,
\label{eq:EMT_PMF_fourier} 
\end{split}
\end{eqnarray}
where $\rho_{\gamma} = \rho_{\gamma,0} a^{-4}$ is energy density of photons. These stress fluctuations can behave as a source of the CMB anisotropies as follows. The first contribution (called passive mode) comes from gravitational interaction via the Einstein equations. In deep radiation dominated era, anisotropic stress fluctuation $\Pi_{Bj}^i$ can enhance superhorizon metric perturbations. After neutrinos decouple, $\Pi_{Bj}^i$ is compensated by anisotropic stress fluctuation of neutrinos and such growth ends. Resulting superhorizon curvature perturbations and gravitational waves are estimated as \cite{Shaw:2009nf} 
\begin{eqnarray}
\begin{split}
\zeta({\bf k}) &=
R_\gamma \ln\left(\frac{\tau_\nu}{\tau_B}\right) 
\frac{3}{2} O_{ij}^{(0)}(\hat{\bf k}) \Pi_{B ij}({\bf k}) ~, 
\\
%----
h^{(\pm 2)}({\bf k}) &= 
6 R_\gamma \ln\left(\frac{\tau_\nu}{\tau_B}\right)  
\frac{1}{2}O_{ij}^{(\mp 2)}(\hat{\bf k}) \Pi_{B ij}({\bf k}) ~,
\label{eq:initial_perturbation}
\end{split}
\end{eqnarray}
where $R_\gamma = 0.6$, $\tau_\nu$, $\tau_B$, $O_{ij}^{(0)}$ and $O_{ij}^{(\pm 2)}$ are ratio of $\rho_\gamma$ divided by total radiation energy density, conformal times of neutrino decoupling and PMF generation, and scalar-mode and tensor-mode projection tensors, respectively \cite{Shiraishi:2012rm}. These re-enter horizon just before recombination and generate CMB scalar and tensor fluctuations. Note that vector-mode metric perturbation decays after neutrino decoupling. These passive-mode anisotropies have similar shapes as the CMB fluctuations in non-magnetized standard cosmology \cite{Pritchard:2004qp} since the changes of metric perturbations mentioned above do not affect radiation transfer functions. The second contribution (called compensated mode) is due to Lorentz force at around recombination. The Lorentz force induces baryon velocity via the Euler equations and enhances the CMB scalar and vector fluctuations \cite{Subramanian:1998fn, Subramanian:2002nh, Subramanian:2003sh, Paoletti:2008ck, Shaw:2009nf}. Unlike the passive mode, the compensated-mode fluctuations are amplified on small scales and hence they differ from standard CMB patterns. From the analyses of these effects by the CMB power spectra, the PMF strength smoothed on 1 Mpc and the spectral index of the PMF power spectrum have been estimated as $B_1 < 3.4 ~{\rm nG}$ and $n_B < 0$ preferred at 95\% CL \cite{Ade:2013lta}.

The tensor and scalar passive modes dominate over the temperature and E-mode fluctuations for $\ell \lesssim 2000$ \cite{Shaw:2009nf, Shiraishi:2012rm}. Even in the B-mode fluctuation, the vector compensated mode is hidden by the presence of the tensor passive mode up to $\ell \sim 500$. In the following discussions, we are interested in scales which are not so small; therefore we shall take into account the effects of the scalar and tensor passive modes.

%//////////////////////////////// 
\subsection{CMB bispectra}

PMF-induced metric perturbations (\ref{eq:initial_perturbation}) obey chi-square statistics because of Gaussianity of PMFs. These induce large squeezed-type curvature and tensor bispectra and will be observed as the temperature and polarization bispectra at the present time. In general, these bispectra have very complicated spin and angle dependence due to contraction of $O_{ij}^{(0)}$, $O_{ij}^{(\pm 2)}$ and the bispectrum of $\Pi_{Bj}^i$ \cite{Shiraishi:2011fi, Shiraishi:2012rm, Shiraishi:2013vja}. In addition, owing to the dependence of the CMB bispectra on $B_i^6$, we are enforced to deal with loop computation. Using a suitable approximation picking up poles, ref.~\cite{Shiraishi:2012rm} has derived complete formulae applicable to the bispectra composed of not only temperature ($I$) but also E-mode ($E$) and B-mode ($B$) anisotropies. 

\begin{figure}[t]
  \begin{tabular}{cc}
    \begin{minipage}{0.5\hsize}
  \begin{center}
    \includegraphics[width=7.5cm,clip]{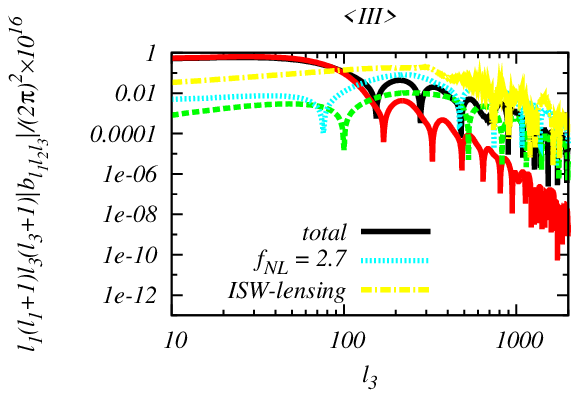}
  \end{center}
\end{minipage}
\begin{minipage}{0.5\hsize}
  \begin{center}
    \includegraphics[width=7.5cm,clip]{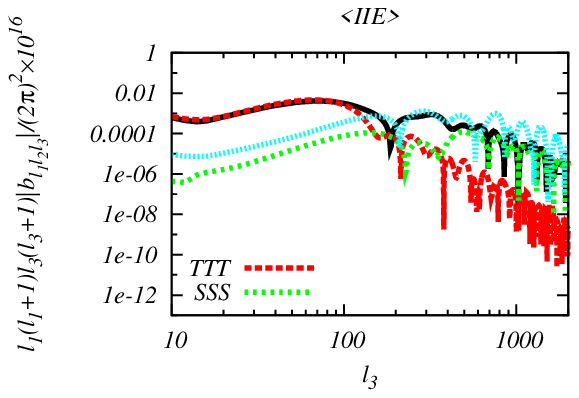}
  \end{center}
\end{minipage}
\end{tabular}
%---------
  \begin{tabular}{cc}
    \begin{minipage}{0.5\hsize}
  \begin{center}
    \includegraphics[width=7.5cm,clip]{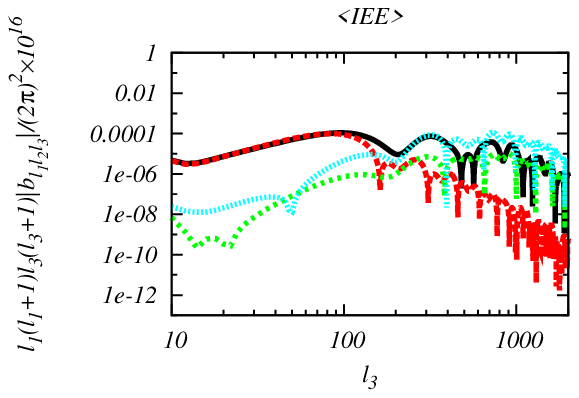}
  \end{center}
\end{minipage}
\begin{minipage}{0.5\hsize}
  \begin{center}
    \includegraphics[width=7.5cm,clip]{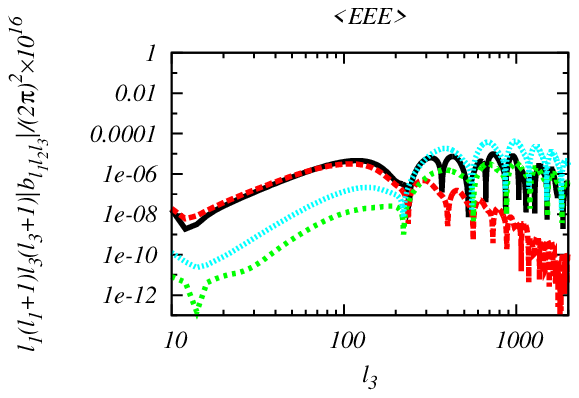}
  \end{center}
\end{minipage}
\end{tabular}
%----------
  \begin{tabular}{cc}
    \begin{minipage}{0.5\hsize}
  \begin{center}
    \includegraphics[width=7.5cm,clip]{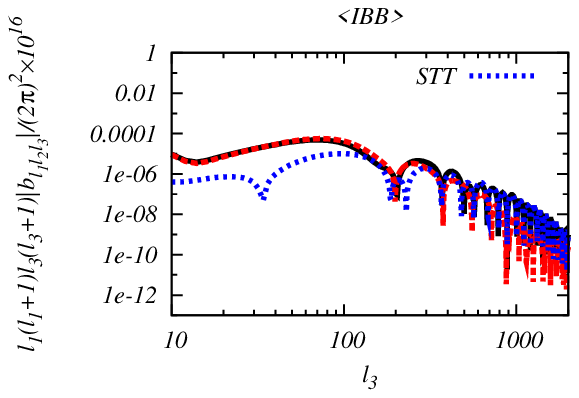}
  \end{center}
\end{minipage}
\begin{minipage}{0.5\hsize}
  \begin{center}
    \includegraphics[width=7.5cm,clip]{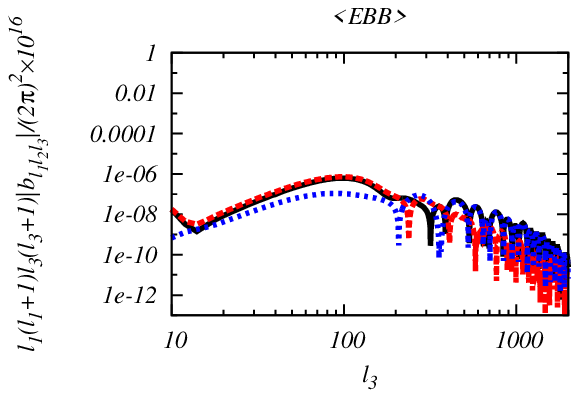}
  \end{center}
\end{minipage}
\end{tabular}
  \caption{Parity-even magnetized bispectra: $\Braket{III}$, $\Braket{IIE}$, $\Braket{IEE}$, $\Braket{EEE}$, $\Braket{IBB}$ and $\Braket{EBB}$ for $\ell_1 = \ell_2 = \ell_3$. For comparison, the local-type bispectra with $f_{\rm NL} = 2.7$ and the ISW-lensing bispectrum are also plotted. The $total$ spectrum means the bispectrum involving all possible scalar and tensor combinations, and $TTT$, $SSS$ or $STT$ corresponds to a part of the $total$ bispectrum. The PMF parameters are taken as $B_{1} = 3 ~{\rm nG}$, $\tau_\nu / \tau_B = 10^{17}$ and $n_B = -2.9$. Other cosmological parameters are fixed as values consistent with the {\it Planck} results \cite{Ade:2013lta}. These spectra obey $\ell_1 + \ell_2 + \ell_3 = {\rm even}$.} \label{fig:bis_even}
\end{figure}

\begin{figure}[t]
  \begin{tabular}{cc}
    \begin{minipage}{0.5\hsize}
  \begin{center}
    \includegraphics[width=7.5cm,clip]{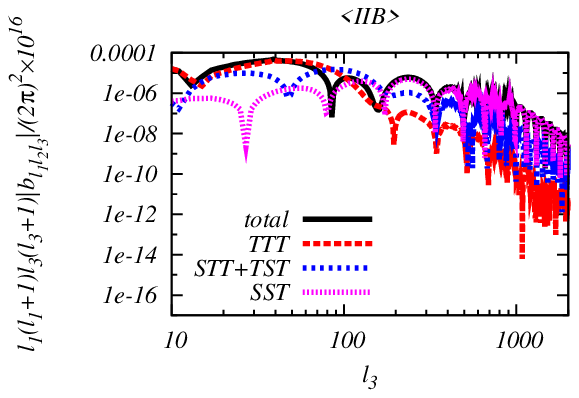}
  \end{center}
\end{minipage}
\begin{minipage}{0.5\hsize}
  \begin{center}
    \includegraphics[width=7.5cm,clip]{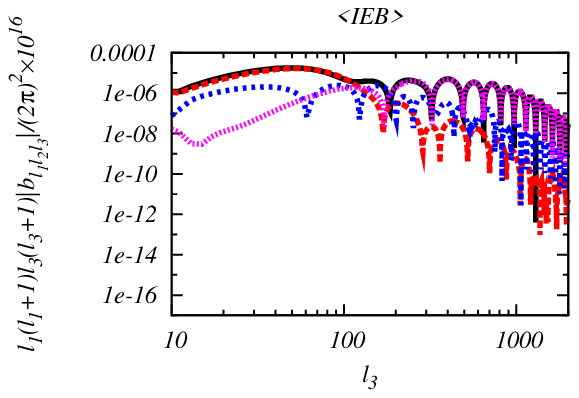}
  \end{center}
\end{minipage}
\end{tabular}
%---------
  \begin{tabular}{cc}
    \begin{minipage}{0.5\hsize}
  \begin{center}
    \includegraphics[width=7.5cm,clip]{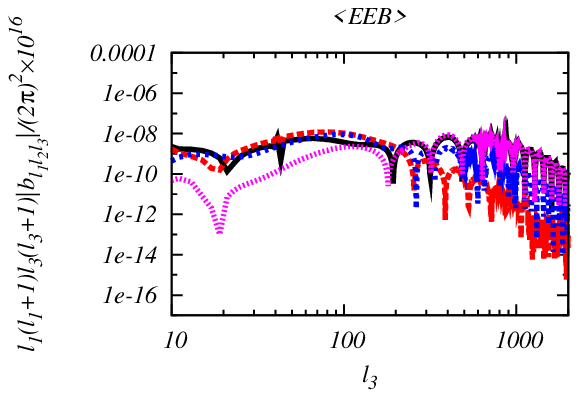}
  \end{center}
\end{minipage}
\begin{minipage}{0.5\hsize}
  \begin{center}
    \includegraphics[width=7.5cm,clip]{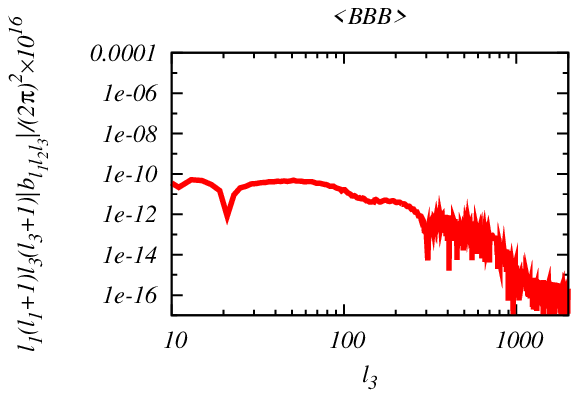}
  \end{center}
\end{minipage}
\end{tabular}
  \caption{Parity-odd magnetized bispectra: $\Braket{IIB}$, $\Braket{IEB}$, $\Braket{EEB}$ and $\Braket{BBB}$ for $\ell_1 + 4 = \ell_2 + 2 = \ell_3$. The $total$ bispectrum consists of the auto- and cross-bispectra of the scalar and tensor anisotropies such as $TTT$, $STT+TST$ and $SSS$. The settings for the PMF parameters and other cosmological parameters are same as figure~\ref{fig:bis_even}. These spectra obey $\ell_1 + \ell_2 + \ell_3 = {\rm odd}$. Rapidly-oscillating behavior seen in $\Braket{BBB}$ is due to antisymmetric property of the parity-odd bispectrum.} \label{fig:bis_odd}
\end{figure}

Computing on the basis of their formalism, we depict the CMB bispectra in figures~\ref{fig:bis_even} and \ref{fig:bis_odd}. Here we distinguish between six parity-even bispectra ($\Braket{III}$, $\Braket{IIE}$, $\Braket{IEE}$, $\Braket{EEE}$, $\Braket{IBB}$ and $\Braket{EBB}$) and four parity-odd ones ($\Braket{IIB}$, $\Braket{IEB}$, $\Braket{EEB}$ and $\Braket{BBB}$) because they are located at completely different multipole configurations, namely $\ell_1 + \ell_2 + \ell_3 = {\rm even}$ and ${\rm odd}$, respectively. The vertical axes express absolute values of reduced bispectra: $b_{\ell_1 \ell_2 \ell_3} = G_{\ell_1 \ell_2 \ell_3}^{-1} B_{\ell_1 \ell_2 \ell_3}$, where 
\begin{eqnarray}
\Braket{\prod_{n=1}^3 a_{\ell_n m_n}} &\equiv&  
\left(
  \begin{array}{ccc}
  \ell_1 & \ell_2 & \ell_3 \\
  m_1 & m_2 & m_3 
  \end{array}
 \right) B_{\ell_1 \ell_2 \ell_3} ~, \\
%-----------
G_{\ell_1 \ell_2 \ell_3} 
&\equiv& \frac{1}{6} \left[ \frac{2 \sqrt{\ell_3 (\ell_3 + 1) \ell_2 (\ell_2 +
1)}}{\ell_1(\ell_1 + 1) - \ell_2 (\ell_2 + 1) - \ell_3 (\ell_3 + 1)} \right. \nonumber \\ 
&&\left. \qquad\times 
\sqrt{\frac{\prod_{n=1}^3 (2 \ell_n + 1)}{4 \pi}}
\left(
  \begin{array}{ccc}
  \ell_1 & \ell_2 & \ell_3 \\
   0 & -1 & 1
  \end{array}
 \right) + {5~\rm perms.} \right].
\end{eqnarray} 
Note that a relation: $G_{\ell_1 \ell_2 \ell_3} = \sqrt{\frac{(2 l_1 + 1)(2 l_2 + 1)(2 l_3 + 1)}{4 \pi}}
\left(
  \begin{array}{ccc}
  \ell_1 & \ell_2 & \ell_3 \\
  0 & 0 & 0
  \end{array}
 \right) $ holds when $\ell_1 + \ell_2 + \ell_3 = {\rm even}$ \cite{Kamionkowski:2010rb, Shiraishi:2011st, Shiraishi:2012sn}. The magnetized bispectrum consists of auto- and cross-correlations between the scalar and tensor anisotropies (i.e., $TTT$, $STT$, $TST$, $TTS$, $SST$, $STS$, $TSS$ and $SSS$). In these figures, we express the magnetized bispectrum composed of every conceivable combination in these eight modes as $total$ mode, which means  
\begin{eqnarray}
total = \begin{cases}
\parbox{5cm}{
 \flushleft 
$TTT + TTS+TST+STT $\\ 
$+ SST+STS+TSS + SSS$} & 
\parbox{6cm}{
 \flushleft
: $\Braket{III}$, $\Braket{IIE}$, $\Braket{IEE}$, $\Braket{EEE}$ } \\ 
%--
\parbox{5cm}{
 \flushleft 
$TTT + TST+STT + SST$} & 
\parbox{6cm}{
 \flushleft 
: $\Braket{IIB}$, $\Braket{IEB}$, $\Braket{EEB}$ } \\ 
%--
\parbox{5cm}{
 \flushleft 
$TTT + STT$ } & 
\parbox{6cm}{
 \flushleft 
: $\Braket{IBB}$, $\Braket{EBB}$ } \\ 
%--
\parbox{5cm}{
 \flushleft 
$TTT$ } & 
\parbox{6cm}{
 \flushleft 
: $\Braket{BBB}$ } 
\end{cases} ~.
\end{eqnarray}
Here the difference in the number of terms by each line is due to a fact that the scalar mode cannot generate the B-mode polarization. We also plot each mode to clarify its contribution to the $total$ spectrum.

From these figures, we can confirm that the tensor mode dominates on large scales and the scalar mode catches up with the tensor mode on small scales. These are consistent behaviors with the magnetized power spectra and temperature auto-bispectrum \cite{Shaw:2009nf, Shiraishi:2012rm}. Especially, we can observe that the $TTT$ modes are ${\cal O}(10^2)$ times larger than the $SSS$ modes on sufficient large scales. This amplification directly reflects a magnitude relationship between magnetized gravitational waves and curvature perturbations of eq.~(\ref{eq:initial_perturbation}), namely, $(h^{(\pm 2)}/\zeta)^3 \sim 6^3$. Overall behaviors of the magnetized bispectra are consistent with the CMB power spectra predicted by the standard cosmology because their transfer functions are same. In four panels for the temperature and E-mode bispectra, the standard local-type bispectra are also plotted. Furthermore, in $\Braket{III}$ panel, we also describe a CMB bispectrum from a correlation between the late-time ISW effect and weak lensing, i.e., the ISW-lensing bispectrum \cite{Smith:2006ud, Hanson:2009kg, Mangilli:2009dr, Lewis:2011fk, Pearson:2012ba, Junk:2012qt, Lewis:2012tc, Hu:2012td, DiValentino:2012yg}. It is well known that the ISW-lensing bispectrum highly correlates with the local-type bispectrum. We can see that these two types of bispectra resemble the magnetized $SSS$ bispectra, while they are quite different from the $total$ spectra because of the tensor-mode contributions. Therefore, the magnetized bispectrum signals will not be biased in a multi-parameter fitting (for details see the next section). In the next section, we evaluate the detectability of these signals. 

%--------

%%%%%%%%%%%%%%%%%%%%%%%%%%%%%%%%%%%%%%%%%%%%%%%%%%%%%%
\section{Fisher forecast}\label{sec:Fisher}

In this section, through the Fisher analysis, we evaluate the expected error bar of the magnitude of the magnetized bispectra for $n_B = -2.9$, which depends on the PMF strength (smoothed on 1 Mpc) and PMF generation epoch:
\begin{eqnarray}
A_{\rm bis} = \left( \frac{B_1}{3 ~\rm nG} \right)^6 
\left[ \frac{\ln \left(\tau_\nu / \tau_B\right)}{\ln(10^{17})} \right]^3 ~.
\end{eqnarray}
We assume noise information of temperature and polarizations in the {\it Planck} and PRISM experiments \cite{:2006uk, Andre:2013afa} (for details see appendix~\ref{appen:noise}).

%//////////////////////////////////
\subsection{Temperature and E-mode bispectra}\label{subsec:I+E}

Here we focus on the parity-even signals arising from $\Braket{III}$, $\Braket{IIE}$, $\Braket{IEE}$ and $\Braket{EEE}$. The Fisher matrix element of the normalized bispectra including E-mode polarizations is defined as \cite{Babich:2004yc, Yadav:2007rk}
\begin{eqnarray}
F_{ij} = \sum_{\substack{ X_1 X_2 X_3 \\ X_1' X_2' X_3'}} \sum_{\ell_1 \leq \ell_2 \leq \ell_3 \leq \ell_{\rm max}} 
\frac{1}{\Delta_{\ell_1 \ell_2 \ell_3}} 
\tilde{B}_{X_1' X_2' X_3', \ell_1 \ell_2 \ell_3}^{(i)} 
\left[\prod_{n=1}^3 (C^{-1})_{\ell_n}^{X_n X_n'} \right]
\tilde{B}_{X_1 X_2 X_3, \ell_1 \ell_2 \ell_3}^{(j)} ~,
\end{eqnarray}
where  
\begin{eqnarray}
\Delta_{\ell_1 \ell_2 \ell_3} = (-1)^{\ell_1 + \ell_2 + \ell_3} 
(1 + 2 \delta_{\ell_1, \ell_2} \delta_{\ell_2, \ell_3})  
+ \delta_{\ell_1, \ell_2} + \delta_{\ell_2, \ell_3} + \delta_{\ell_3, \ell_1}~, 
\end{eqnarray}
and $X_1 X_2 X_3$ and $X_1' X_2' X_3'$ run over eight modes $III$, $IIE$, $IEI$, $EII$, $IEE$, $EIE$, $EEI$ and $EEE$. The inverse matrix of the power spectrum is explicitly written as 
\begin{eqnarray}
(C^{-1})_{\ell}^{X X'} \equiv 
\left(
  \begin{array}{cc}
  C_{\ell}^{II} & C_{\ell}^{IE}  \\
  C_{\ell}^{EI} & C_{\ell}^{EE}
  \end{array}
 \right)^{-1} ~,
\end{eqnarray} 
where $C_{\ell}^{XX'} = \bar{C}_{\ell}^{XX'} + N_{\ell}^{XX'}$ is the CMB power spectrum involving information of cosmic variance $\bar{C}_{\ell}$ and instrumental noise $N_{\ell}$. We want to estimate the signals of the magnetized bispectrum ($B^{({M})}$) under the contamination of the local-type bispectrum ($B^{({L})}$) or the ISW-lensing bispectrum ($B^{({\phi})}$) and accordingly $\tilde{B}^{(i, j)} = B^{({M})}/A_{\rm bis}, B^{({L})}/f_{\rm NL}, B^{({\phi})}$.

\begin{figure}[t]
  \begin{center}
    \includegraphics[width=12cm,clip]{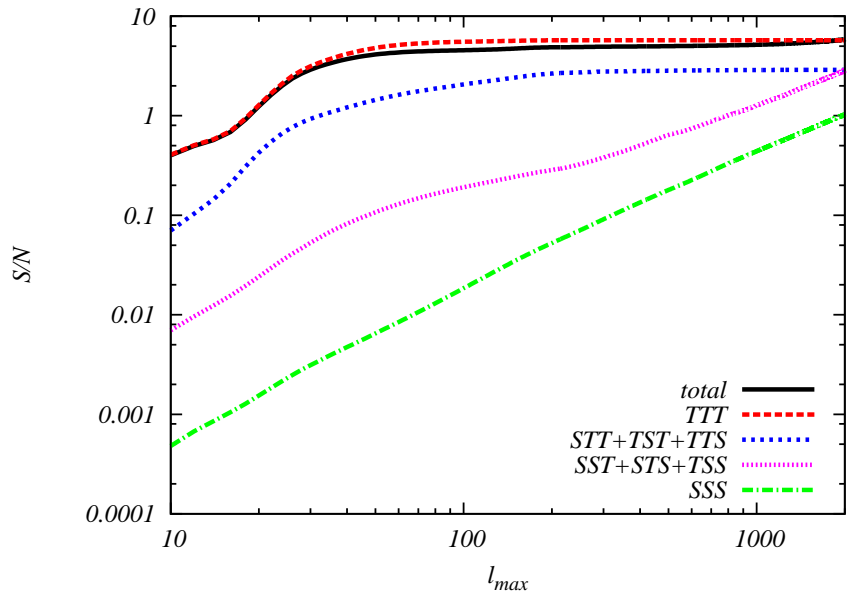}
  \end{center}
  \caption{Signal-to-noise ratios of the magnetized bispectra (\ref{eq:SN_bis}): $\Braket{III} + \Braket{IIE} + \Braket{IEE} + \Braket{EEE}$ when $A_{\rm bis} = 1$. We neglect any instrumental noises; hence the signal-to-noise ratios are determined by the cosmic variance alone.} \label{fig:SN_I+E_ideal}
\end{figure}

Firstly, let us clarify the dependence of the magnetized temperature and polarization bispectra on the scalar and tensor modes under the cosmic-variance-limited ideal experiment. In figure~\ref{fig:SN_I+E_ideal} we plot the signal-to-noise ratio, which is given as  
\begin{eqnarray}
\frac{S}{N} = \sqrt{F_{MM}}~. \label{eq:SN_bis}
\end{eqnarray}
From this figure, we can see that contribution of the tensor mode is quite larger than that of the scalar mode and therefore the $TTT$ mode dominates over the $total$ spectrum. However, due to rapidly decaying nature, the tensor mode is saturated for $\ell \gtrsim 100$ and the scalar mode also contributes to a bit of amplification of the $total$ spectrum. Note that the $total$ spectrum falls below the $TTT$ mode due to sign difference of each mode. These features have also been observed in the analysis of $\Braket{III}$ \cite{Shiraishi:2012rm} and are quite different from the local-type bispectrum signatures that behave as simple increasing functions of $\ell_{\rm max}$ \cite{Babich:2004yc}. 

Next, to estimate the uncertainty of $A_{\rm bis}$ under the presence of the contamination of the local-type bispectrum, we introduce the Fisher submatrix as 
\begin{eqnarray}
{}^{(2)}F &=& \left(
  \begin{array}{cc}
  F_{MM} & F_{ML}  \\
  F_{LM} & F_{LL}
  \end{array}
 \right)~.
\end{eqnarray} 
Then, the $1\sigma$ errors are given by 
\begin{eqnarray}
(\delta A_{\rm bis}, \delta f_{\rm NL}) 
= \left( \sqrt{{}^{(2)}F^{-1}_{11}}, \sqrt{{}^{(2)}F^{-1}_{22}} \right) ~. \label{eq:Abis_fnl}
\end{eqnarray}
Numerical results of $\delta A_{\rm bis}$ are described in figure~\ref{fig:error_I+E}. We will also present $\delta f_{\rm NL}$ in appendix~\ref{appen:noise}. From this figure, it is found that if we use all information of the temperature and E-mode bispectra, $\delta A_{\rm bis}$ is $80\%$ improved in comparison with the analysis in terms of $\Braket{III}$ alone under the ideal case. This is an interesting result since in estimation for the local-type bispectrum $\delta f_{\rm NL}$ is only $50\%$ reduced (See refs.~\cite{Babich:2004yc, Yadav:2007rk} or figure~\ref{fig:error_fnl}). This indicates that the tensor-mode polarization bispectra are quite informative. As described in this figure, measuring $A_{\rm bis}$ with this accuracy is hard in the {\it Planck} experiment due to lack of sensitivity of polarizations (see appendix~\ref{appen:noise}), while it can be done in the PRISM experiment. In the {\it Planck}, PRISM and ideal experiments for $\ell_{\rm max} = 2000$, we obtain $\delta A_{\rm bis} = 0.46$, $0.17$ and $0.17$, respectively (table~\ref{tab:Abis}). 

To quantify resemblance between $B^{(M)}$ and $B^{(L)}$, we may compute a shape correlator given by 
\begin{eqnarray}
r_{ML} \equiv \frac{F_{ML}}{\sqrt{F_{MM} F_{LL}}} ~.
\end{eqnarray}
A numerical result for $\ell_{\rm max } = 2000$, i.e., $r_{M L} = -0.17$, guarantees that the magnetized bispectrum is weakly correlated with the local-type bispectrum and its contamination is very small. As this result, $(S/N)^{-1}$ of the $total$ spectrum in figure~\ref{fig:SN_I+E_ideal} coincides with $\delta A_{\rm bis} = 0.17$. 

\begin{figure}[t]
  \begin{center}
    \includegraphics[width=12cm,clip]{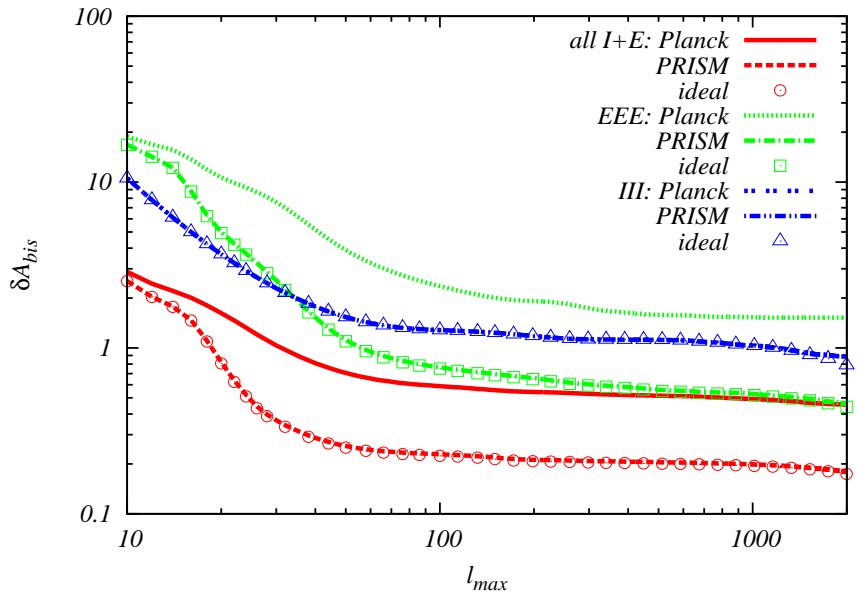}
  \end{center}
  \caption{Expected $1\sigma$ errors of $A_{\rm bis}$ (\ref{eq:Abis_fnl}) estimated from all possible temperature and E-mode bispectra (red), only $\Braket{EEE}$ (green) and only $\Braket{III}$ (blue) if we assume the {\it Planck}, PRISM and ideal noise spectra.} \label{fig:error_I+E}
\end{figure} 

Finally, let us evaluate the bias by the ISW-lensing bispectrum. This contaminates only $\Braket{III}$. In the same manner as the above discussion, we compute $\delta A_{\rm bis}$ by following 
\begin{eqnarray}
\delta A_{\rm bis} &=& \sqrt{ {}^{(2)}F_{11}'^{-1} } ~, \\
{}^{(2)}F' &=& \left(
  \begin{array}{cc}
  F_{MM} & F_{M \phi}  \\
  F_{\phi M} & F_{\phi \phi}
  \end{array}
 \right) ~,
\end{eqnarray}
and find that the values for $\ell_{\rm max} = 2000$ become $0.92$ ({\it Planck}), $0.92$ (PRISM) and $0.83$ (ideal), respectively. These are almost identical to the values of $\delta A_{\rm bis}$ from $\Braket{III}$ in figure~\ref{fig:error_I+E} (or table~\ref{tab:Abis}) and hence we can conclude that the ISW-lensing bispectrum is also a tiny bias comparable to the local-type bispectrum in the $\Braket{III}$ analysis.

%////////////////////////////////// 
\subsection{B-mode bispectra}

In this subsection, we shall consider a possibility of the bispectra including B-mode polarization. Such bispectra are divided into both the parity-even ($\Braket{IBB}$ and $\Braket{EBB}$) and the parity-odd ($\Braket{IIB}$, $\Braket{IEB}$, $\Braket{EEB}$ and $\Braket{BBB}$) combinations. Although a complete analysis with both these all contributions and the temperature and E-mode bispectra may reduce $\delta A_{\rm bis}$ more drastically, it will be quite complicated. Accordingly, here let us concentrate on the Fisher analysis with $\Braket{BBB}$ alone. 

For $\ell \gtrsim 500$, the compensated vector mode will exceeds the passive tensor mode. Furthermore, on such scales, lensed CMB fluctuations also generate secondary B-mode fluctuations and may contaminate the magnetized bispectrum \cite{Shaw:2009nf, Lewis:2011fk, Hanson:2013hsb}. While the consideration of these sources is important, in this paper we work on large scales up to $\ell_{\rm max} = 500$ where these are negligible.

\begin{figure}[t]
  \begin{center}
    \includegraphics[width=12cm,clip]{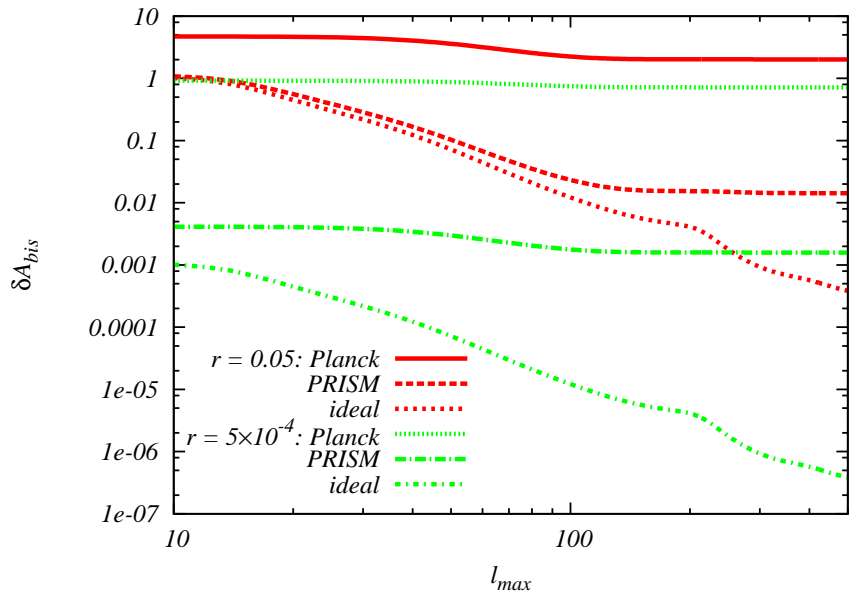}
  \end{center}
  \caption{Expected $1\sigma$ errors of $A_{\rm bis}$ (\ref{eq:Abis_BBB}) estimated from $\Braket{BBB}$ if we assume the {\it Planck}, PRISM or ideal noise spectrum with $r = 0.05$ or $5\times 10^{-4}$.} \label{fig:error_BBB}
\end{figure}

Despite the parity-odd case, we can define the Fisher matrix like the parity-even case:
\begin{eqnarray}
F \equiv \sum_{\ell_1 \leq \ell_2 \leq \ell_3 \leq {\ell_{\rm max}}} 
\frac{ \tilde{B}_{BBB, \ell_1 \ell_2 \ell_3}^2}{\Delta_{\ell_1 \ell_2 \ell_3} 
\prod_{n=1}^3 C^{BB}_{\ell_n}} ~, 
\end{eqnarray}
where $\tilde{B} = B^{(M)} / A_{\rm bis}$. Then the $1\sigma$ error becomes 
\begin{eqnarray}
\delta A_{\rm bis} = \sqrt{F^{-1}}~. \label{eq:Abis_BBB}
\end{eqnarray}
Figure~\ref{fig:error_BBB} describes the numerical results of $\delta A_{\rm bis}$. As the cosmic-variance spectrum $\bar{C}_{\ell}^{BB}$, we adopt non-magnetized tensor-mode power spectrum in the standard cosmology, whose amplitude is determined by the tensor-to-scalar ratio $r$. Especially for the ideal case ($N_{\ell}^{BB} = 0$), $\delta A_{\rm bis}$ is then simply proportional to $r^{3/2}$ and therefore we can write $\delta A_{\rm bis} \approx 0.03 r^{3/2}$ for $\ell_{\rm max} = 500$. Interestingly, unlike the estimation with the temperature and E-mode bispectra, $\delta A_{\rm bis}$ in the ideal experiment does not saturate even for high $\ell_{\rm max}$. This is due to damping behavior of $\bar{C}_{\ell}^{BB}$ for $\ell \gtrsim 100$, which cannot be seen in $\bar{C}_{\ell}^{II}, \bar{C}_{\ell}^{IE}$ and $\bar{C}_{\ell}^{EE}$ (see figure~\ref{fig:noise}). For $r = 0.05$, owing to this effect, $\delta A_{\rm bis}$ reaches $0.014$ under the PRISM noise level. On the other hand, the {\it Planck} experiment is too noisy to reduce the error so much like the $\Braket{EEE}$ case. If $r = 5 \times 10^{-4}$, the noise dominates completely and $\delta A_{\rm bis}$ saturates for all $\ell_{\rm max}$ in both the experiments. 

Finally, we summarize the value of $\delta A_{\rm bis}$ for each case in table~\ref{tab:Abis}

\begin{table}[t]
\begin{center}
  \begin{tabular}{|c||c|c|c|c|c|} \hline
    experiment & $III$ & $EEE$ & all $I+E$ & $BBB$ $(r = 0.05)$ & $BBB$ $(r = 5\times 10^{-4})$ \\ \hline 
    {\it Planck} & 0.89 & 1.5 & 0.46 & 2.0 & 0.71 \\ 
    PRISM & 0.89 & 0.46 & 0.17 & $1.4 \times 10^{-2}$ & $1.6 \times 10^{-3}$  \\
    ideal & 0.79 & 0.44 & 0.17 & $3.4 \times 10^{-4}$ & $3.4 \times 10^{-7}$ \\ \hline
  \end{tabular}
\end{center}
\caption{Expected $1\sigma$ errors of $A_{\rm bis}$ at $\ell_{\rm max} = 2000$ ($III$, $EEE$ and all $I+E$) and $500$ ($BBB$) for each experiment.}\label{tab:Abis}
\end{table}

%%%%%%%%%%%%%%%%%%%%%%%%%%%%%%%%%%%%%%%%%%%%%%%%%%%%%%
\section{Summary and discussion}\label{sec:summary}

In this paper we examined how the polarization bispectra of the CMB anisotropies affect constraining PMFs. Firstly, we confirmed that the tensor-mode signals dominate over the bispectrum for $\ell \gtrsim 2000$ and the scalar mode contributes on very small scales in the auto- and cross-bispectra with the polarizations. Owing to this dependence, the magnetized bispectra are weakly correlated with the standard local-type bispectra and the ISW-lensing bispectrum, and hence from observations the information of PMFs will be able to be extracted efficiently without any contamination. 

From the error analyses via the Fisher forecast, we found that potentially, if we utilize all the temperature and E-mode bispectra, the uncertainty of the magnitude of magnetized bispectra can be $80\%$ improved in comparison with the analysis with respect to $\Braket{III}$ alone. This is interesting since in the analysis of the local-type non-Gaussianity, the improvement is only $50\%$. The proposed PRISM experiment will be able to reach this precision, while the {\it Planck} experiment cannot. If we assume the GUT-scale generation of PMFs, namely $\tau_\nu / \tau_B = 10^{17}$, the expected $1\sigma$ errors on the PMF strength from all the temperature and E-mode bispectra are given as $\delta B_1 / {\rm nG} = 2.6$ and $2.2$ in the {\it Planck} and PRISM (or ideal) experiments, respectively. 

We also considered the possibility of the analysis involving the B-mode bispectrum. In this case, we focused on the Fisher forecast using $\Braket{BBB}$ and found that the uncertainty keeps on reducing as $\ell_{\rm max}$ increases due to the damping behavior of the B-mode cosmic-variance spectrum for $\ell \gtrsim 100$. In the ideal experiment, we have a relationship with the tensor-to-scalar ratio: $\delta B_1 / {\rm nG} \approx 1.7 r ^{1/4}$ for $\ell_{\rm max} = 500$; therefore we will be able to estimate with ${\cal O}(0.1)$ nG accuracy if $r \lesssim 0.1$. In practice, the {\it Planck} and PRISM instrumental noises relax the value as $\delta B_1 / {\rm nG} = 3.4 ~(2.8)$ and $1.5~ (1.0)$ for $r = 0.05~ (5 \times 10^{-4})$, respectively.

One may be concerned about comparison with bounds from the power spectrum analysis. According to recent literature \cite{Paoletti:2010rx, Shaw:2010ea, Paoletti:2012bb, Ade:2013lta}, upper bounds on $B_1$ from the temperature and E-mode power spectra are around 3 nG. As shown above, the bispectrum analysis will provide comparable or tighter constraints on $B_1$. Concerning the B-mode power spectrum, the magnetized passive-mode signals are indistinguishable from the non-magnetized ones from primordial gravitational waves and hence $B_1$ may be not determined accurately in a multi-parameter fitting. In this sense, the information of the B-mode bispectrum will be more useful. 

For $\ell > 500$, where this paper has not focused on, the vector compensated mode will dominate over the magnetized B-mode bispectrum. To reduce the uncertainty, we must evaluate such vector-mode contribution. Then, more comprehensive analysis including cross-bispectra between temperature, E-mode and B-mode fluctuations will be required. These informative but complex works remain as future issues. 

%%%%%%%%%%%%%%%%%%%%%%%
\acknowledgments
We thank Sabino Matarrese for helpful advice and motivational comments. We appreciate helpful advice on estimation of noise spectra given by Bin Hu and Michele Liguori. This work was supported in part by a Grant-in-Aid for JSPS Research under Grant No.~25-573 and the ASI/INAF Agreement I/072/09/0 for the Planck LFI Activity of Phase E2.

\appendix
%%%%%%%%%%%%%%%%%%%%%%%%%%%%%%%%%%%%%%%%%%%%%%%%%%%%%
\section{Noise spectra}\label{appen:noise}

Here, we summarize the temperature and polarization noise spectra expected in the {\it Planck} and PRISM experiments. 

Assuming Gaussian random detector noise, each noise spectrum is estimated as \cite{Hu:2001fb, Hu:2012td, Santos:2013gqa} 
\begin{eqnarray}
N_l^{XX} = \left[ \sum_c \frac{1}{\theta_c^2 \sigma_{X, c}^2} e^{-\ell(\ell+1)\theta_c^2 / (8\ln 2)}  \right]^{-1} \label{eq:Nl}
\end{eqnarray}
where $\theta_c$ is the Full Width Half Maximum (FWHM) per the frequency channel $c$ in radians and $\sigma_{X,c}$ is the dimensionless sensitivity per $c$. One can find these values (in arcminutes and $\mu$K) in table~{\ref{tab:noise}}. 

\begin{figure}[t]
  \begin{center}
    \includegraphics[width=12cm,clip]{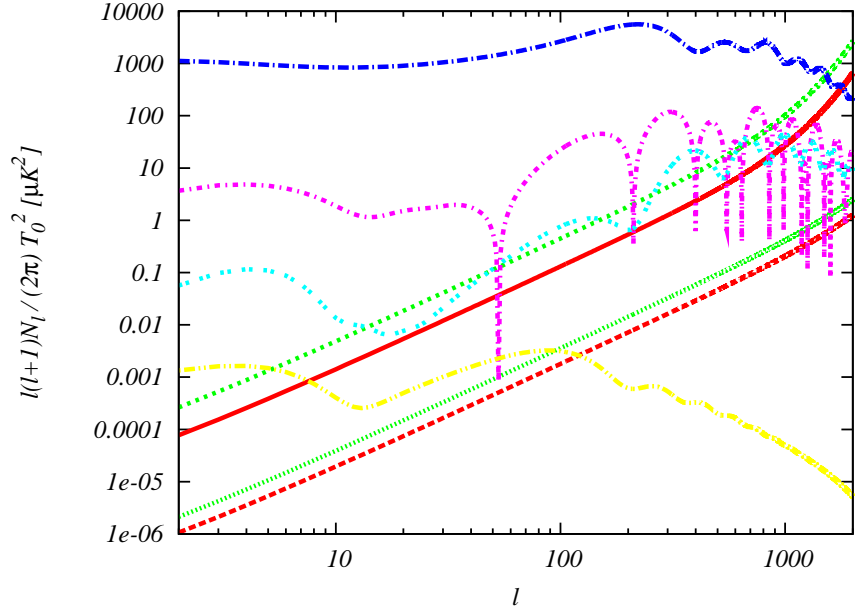}
  \end{center}
  \caption{Noise spectra $N_\ell^{II}$ (red) and $N_\ell^{EE/BB}$ (green) (\ref{eq:Nl}) assuming the {\it Planck} (top) and PRISM (bottom) experiments. For comparison, we also depict $\bar{C}_\ell^{II}$ (blue), $|\bar{C}_{\ell}^{IE}|$ (magenta), $\bar{C}_{\ell}^{EE}$ (cyan) and $\bar{C}_{\ell}^{BB}$ for $r = 0.05$ (yellow).} \label{fig:noise}
\end{figure}

Figure~\ref{fig:noise} shows numerical results of $N_\ell^{II}$, $N_{\ell}^{EE}$ and $N_{\ell}^{BB}$. Here we assume $N_\ell^{IE} = 0$. We can see that in the $EE$ mode, the cosmic-variance spectrum is comparable to the {\it Planck} noise spectrum, i.e., $\bar{C}_\ell^{EE} \sim N_\ell^{EE}$, for $\ell \gtrsim 10$. This is a reason why the {\it Planck} experiment does not improve $\delta A_{\rm bis}$ so much in the analysis including the $E$-mode polarization as described in figure~\ref{fig:error_I+E}. Likewise, in the PRISM experiment $N_\ell^{BB}$ exceeds $\bar{C}_\ell^{BB} (r = 0.05)$ for $\ell \gtrsim 100$ and hence $\delta A_{\rm bis}$ never be reduced beyond $\ell \sim 100$ when $r = 0.05$ (figure~\ref{fig:error_BBB}).

\begin{table}[t]
\begin{center}
  \begin{tabular}{|c||c|c|c|} \hline
    frequency (GHz) & $\theta$ (arcmin) & $\sigma_I$ ($\mu$K) & $\sigma_{E/B}$ ($\mu$K) \\ \hline 
    100 & 9.5 & 6.8 & 10.9 \\  
    143 & 7.1 & 6.0 & 11.5 \\ 
    217 & 5.0 & 13.1 & 26.8 \\
    353 & 5.0 & 40.1 & 81.3 \\ \hline\hline
%-----------
    105 & 4.8 & 2.9 & 4.1 \\
    135 & 3.8 & 2.6 & 3.7 \\  
    160 & 3.2 & 2.4 & 3.4 \\ 
    185 & 2.8 & 2.5 & 3.6 \\
    200 & 2.5 & 2.6 & 3.7 \\ \hline
  \end{tabular}
\end{center}
\caption{Instrumental information of the {\it Planck} (top) and PRISM (bottom) experiments \cite{:2006uk,Andre:2013afa}.} \label{tab:noise}
\end{table}

%%%%%%%%%%%%%%%%%%%%%%%%%%%%%%%%%%%%
\section{Errors of the local-type non-Gaussianity}\label{appen:fnl}

\begin{figure}[t]
  \begin{center}
    \includegraphics[width=12cm,clip]{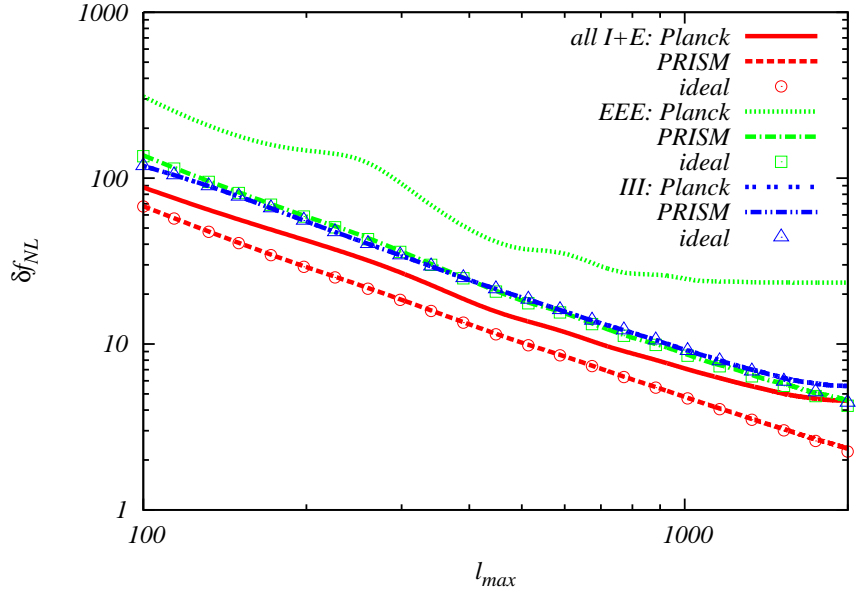}
  \end{center}
  \caption{Expected $1\sigma$ errors of $f_{\rm NL}$ (\ref{eq:Abis_fnl}) estimated from all possible temperature and E-mode bispectra (red), only $\Braket{EEE}$ (green) and only $\Braket{III}$ (blue) if we assume the {\it Planck}, PRISM and ideal noise spectra.} \label{fig:error_fnl}
\end{figure}

In figure~\ref{fig:error_fnl}, we describe the $1\sigma$ errors of the local-type nonlinearity parameter $f_{\rm NL}$ estimated from the two-dimensional Fisher analysis involving $\delta A_{\rm bis}$ discussed in subsection~\ref{subsec:I+E}. Thanks to the weak correlation with the magnetized bispectrum, $\delta f_{\rm NL}$ is in good agreement with the results from the one-dimensional Fisher analysis fitting $f_{\rm NL}$ alone, i.e., $1/\sqrt{F_{LL}}$ \cite{Yadav:2007rk}. We confirm that in the ideal experiment, the analysis containing both the temperature and E-mode bispectra reduces the value of $\delta f_{\rm NL}$ to half in comparison with the analysis by $\Braket{III}$ alone. In the PRISM experiment, $\delta f_{\rm NL}$ can reach $2.3$ for $\ell_{\rm max} = 2000$. 

%%%%%%%%%%%%%%%%%%%%%%%%%%%%%%%%%%%%%%%%
\bibliography{paper}

\providecommand{\href}[2]{#2}\begingroup\raggedright\begin{thebibliography}{10}

\bibitem{Bernet:2008qp}
M.~L. Bernet, F.~Miniati, S.~J. Lilly, P.~P. Kronberg, and
  M.~Dessauges-Zavadsky, {\it {Strong magnetic fields in normal galaxies at
  high redshifts}},  {\em Nature} {\bf 454} (2008) 302--304,
  [\href{http://xxx.lanl.gov/abs/0807.3347}{{\tt arXiv:0807.3347}}].

\bibitem{Wolfe:2008nk}
A.~M. Wolfe, R.~A. Jorgenson, T.~Robishaw, C.~Heiles, and J.~X. Prochaska, {\it
  {An 84 microGauss Magnetic Field in a Galaxy at Redshift z=0.692}},  {\em
  Nature} {\bf 455} (2008) 638, [\href{http://xxx.lanl.gov/abs/0811.2408}{{\tt
  arXiv:0811.2408}}].

\bibitem{Neronov:1900zz}
A.~Neronov and I.~Vovk, {\it {Evidence for strong extragalactic magnetic fields
  from Fermi observations of TeV blazars}},  {\em Science} {\bf 328} (2010)
  73--75, [\href{http://xxx.lanl.gov/abs/1006.3504}{{\tt arXiv:1006.3504}}].

\bibitem{Tavecchio:2010mk}
F.~Tavecchio, G.~Ghisellini, L.~Foschini, G.~Bonnoli, G.~Ghirlanda, {\em
  et.~al.}, {\it {The intergalactic magnetic field constrained by Fermi/LAT
  observations of the TeV blazar 1ES 0229+200}},  {\em Mon.Not.Roy.Astron.Soc.}
  {\bf 406} (2010) L70--L74, [\href{http://xxx.lanl.gov/abs/1004.1329}{{\tt
  arXiv:1004.1329}}].

\bibitem{Dolag:2010ni}
K.~Dolag, M.~Kachelriess, S.~Ostapchenko, and R.~Tomas, {\it {Lower limit on
  the strength and filling factor of extragalactic magnetic fields}},  {\em
  Astrophys.J.} {\bf 727} (2011) L4,
  [\href{http://xxx.lanl.gov/abs/1009.1782}{{\tt arXiv:1009.1782}}].

\bibitem{Takahashi:2013uoa}
K.~Takahashi, M.~Mori, K.~Ichiki, S.~Inoue, and H.~Takami, {\it {Lower Bounds
  on Magnetic Fields in Intergalactic Voids from Long-term GeV-TeV Light Curves
  of the Blazar Mrk 421}},  {\em Astrophys.J.} {\bf 771} (2013) L42.

\bibitem{Bamba:2006ga}
K.~Bamba and M.~Sasaki, {\it {Large-scale magnetic fields in the inflationary
  universe}},  {\em JCAP} {\bf 0702} (2007) 030,
  [\href{http://xxx.lanl.gov/abs/astro-ph/0611701}{{\tt astro-ph/0611701}}].

\bibitem{Martin:2007ue}
J.~Martin and J.~Yokoyama, {\it {Generation of Large-Scale Magnetic Fields in
  Single-Field Inflation}},  {\em JCAP} {\bf 0801} (2008) 025,
  [\href{http://xxx.lanl.gov/abs/0711.4307}{{\tt arXiv:0711.4307}}].

\bibitem{Kanno:2009ei}
S.~Kanno, J.~Soda, and M.-a. Watanabe, {\it {Cosmological Magnetic Fields from
  Inflation and Backreaction}},  {\em JCAP} {\bf 0912} (2009) 009,
  [\href{http://xxx.lanl.gov/abs/0908.3509}{{\tt arXiv:0908.3509}}].

\bibitem{Demozzi:2009fu}
V.~Demozzi, V.~Mukhanov, and H.~Rubinstein, {\it {Magnetic fields from
  inflation?}},  {\em JCAP} {\bf 0908} (2009) 025,
  [\href{http://xxx.lanl.gov/abs/0907.1030}{{\tt arXiv:0907.1030}}].

\bibitem{Ichiki:2006cd}
K.~Ichiki, K.~Takahashi, H.~Ohno, H.~Hanayama, and N.~Sugiyama, {\it
  {Cosmological Magnetic Field: a fossil of density perturbations in the early
  universe}},  {\em Science} {\bf 311} (2006) 827--829,
  [\href{http://xxx.lanl.gov/abs/astro-ph/0603631}{{\tt astro-ph/0603631}}].

\bibitem{Maeda:2008dv}
S.~Maeda, S.~Kitagawa, T.~Kobayashi, and T.~Shiromizu, {\it {Primordial
  magnetic fields from second-order cosmological perturbations:Tight coupling
  approximation}},  {\em Class. Quant. Grav.} {\bf 26} (2009) 135014,
  [\href{http://xxx.lanl.gov/abs/0805.0169}{{\tt arXiv:0805.0169}}].

\bibitem{Fenu:2010kh}
E.~Fenu, C.~Pitrou, and R.~Maartens, {\it {The seed magnetic field generated
  during recombination}},  {\em Mon.Not.Roy.Astron.Soc.} {\bf 414} (2011)
  2354--2366, [\href{http://xxx.lanl.gov/abs/1012.2958}{{\tt
  arXiv:1012.2958}}].

\bibitem{Hollenstein:2012mb}
L.~Hollenstein, R.~K. Jain, and F.~R. Urban, {\it {Cosmological Ohm's law and
  dynamics of non-minimal electromagnetism}},  {\em JCAP} {\bf 1301} (2013)
  013, [\href{http://xxx.lanl.gov/abs/1208.6547}{{\tt arXiv:1208.6547}}].

\bibitem{Saga:2013glg}
S.~Saga, M.~Shiraishi, K.~Ichiki, and N.~Sugiyama, {\it {Generation of magnetic
  fields in Einstein-Aether gravity}},  {\em Phys. Rev. D 87,} {\bf 104025}
  (2013) [\href{http://xxx.lanl.gov/abs/1302.4189}{{\tt arXiv:1302.4189}}].

\bibitem{Suyama:2012wh}
T.~Suyama and J.~Yokoyama, {\it {Metric perturbation from inflationary magnetic
  field and generic bound on inflation models}},  {\em Phys.Rev.} {\bf D86}
  (2012) 023512, [\href{http://xxx.lanl.gov/abs/1204.3976}{{\tt
  arXiv:1204.3976}}].

\bibitem{Demozzi:2012wh}
V.~Demozzi and C.~Ringeval, {\it {Reheating constraints in inflationary
  magnetogenesis}},  {\em JCAP} {\bf 1205} (2012) 009,
  [\href{http://xxx.lanl.gov/abs/1202.3022}{{\tt arXiv:1202.3022}}].

\bibitem{Fujita:2012rb}
T.~Fujita and S.~Mukohyama, {\it {Universal upper limit on inflation energy
  scale from cosmic magnetic field}},  {\em JCAP} {\bf 1210} (2012) 034,
  [\href{http://xxx.lanl.gov/abs/1205.5031}{{\tt arXiv:1205.5031}}].

\bibitem{Ringeval:2013hfa}
C.~Ringeval, T.~Suyama, and J.~Yokoyama, {\it {Magneto-reheating constraints
  from curvature perturbations}},
  \href{http://xxx.lanl.gov/abs/1302.6013}{{\tt arXiv:1302.6013}}.

\bibitem{Ferreira:2013sqa}
R.~J.~Z. Ferreira, R.~K. Jain, and M.~S. Sloth, {\it {Inflationary
  Magnetogenesis without the Strong Coupling Problem}},
  \href{http://xxx.lanl.gov/abs/1305.7151}{{\tt arXiv:1305.7151}}.

\bibitem{Seery:2008ms}
D.~Seery, {\it {Magnetogenesis and the primordial non-gaussianity}},  {\em
  JCAP} {\bf 0908} (2009) 018, [\href{http://xxx.lanl.gov/abs/0810.1617}{{\tt
  arXiv:0810.1617}}].

\bibitem{Caldwell:2011ra}
R.~R. Caldwell, L.~Motta, and M.~Kamionkowski, {\it {Correlation of
  inflation-produced magnetic fields with scalar fluctuations}},  {\em
  Phys.Rev.} {\bf D84} (2011) 123525,
  [\href{http://xxx.lanl.gov/abs/1109.4415}{{\tt arXiv:1109.4415}}].

\bibitem{Urban:2012ib}
F.~R. Urban and T.~K. Koivisto, {\it {Perturbations and non-Gaussianities in
  three-form inflationary magnetogenesis}},  {\em JCAP} {\bf 1209} (2012) 025,
  [\href{http://xxx.lanl.gov/abs/1207.7328}{{\tt arXiv:1207.7328}}].

\bibitem{Barnaby:2012tk}
N.~Barnaby, R.~Namba, and M.~Peloso, {\it {Observable non-gaussianity from
  gauge field production in slow roll inflation, and a challenging connection
  with magnetogenesis}},  {\em Phys.Rev.} {\bf D85} (2012) 123523,
  [\href{http://xxx.lanl.gov/abs/1202.1469}{{\tt arXiv:1202.1469}}].

\bibitem{Motta:2012rn}
L.~Motta and R.~R. Caldwell, {\it {Non-Gaussian features of primordial magnetic
  fields in power-law inflation}},  {\em Phys.Rev.} {\bf D85} (2012) 103532,
  [\href{http://xxx.lanl.gov/abs/1203.1033}{{\tt arXiv:1203.1033}}].

\bibitem{Jain:2012ga}
R.~K. Jain and M.~S. Sloth, {\it {Consistency relation for cosmic magnetic
  fields}},  {\em Phys.Rev.} {\bf D86} (2012) 123528,
  [\href{http://xxx.lanl.gov/abs/1207.4187}{{\tt arXiv:1207.4187}}].

\bibitem{Shiraishi:2012xt}
M.~Shiraishi, S.~Saga, and S.~Yokoyama, {\it {CMB power spectra induced by
  primordial cross-bispectra between metric perturbations and vector fields}},
  {\em JCAP} {\bf 1211} (2012) 046,
  [\href{http://xxx.lanl.gov/abs/1209.3384}{{\tt arXiv:1209.3384}}].

\bibitem{Jain:2012vm}
R.~K. Jain and M.~S. Sloth, {\it {On the non-Gaussian correlation of the
  primordial curvature perturbation with vector fields}},  {\em JCAP} {\bf
  1302} (2013) 003, [\href{http://xxx.lanl.gov/abs/1210.3461}{{\tt
  arXiv:1210.3461}}].

\bibitem{Bartolo:2012sd}
N.~Bartolo, S.~Matarrese, M.~Peloso, and A.~Ricciardone, {\it {The anisotropic
  power spectrum and bispectrum in the $f(\phi) F^2$ mechanism}},  {\em
  Phys.Rev.} {\bf D87} (2013) 023504,
  [\href{http://xxx.lanl.gov/abs/1210.3257}{{\tt arXiv:1210.3257}}].

\bibitem{Shiraishi:2013sv}
M.~{Shiraishi}, S.~{Yokoyama}, K.~{Ichiki}, and T.~{Matsubara}, {\it
  {Scale-dependent bias due to primordial vector fields}},  {\em
  Mon.Not.Roy.Astron.Soc.} {\bf 432} (July, 2013) 2331--2338,
  [\href{http://xxx.lanl.gov/abs/1301.2778}{{\tt arXiv:1301.2778}}].

\bibitem{Shiraishi:2013vja}
M.~Shiraishi, E.~Komatsu, M.~Peloso, and N.~Barnaby, {\it {Signatures of
  anisotropic sources in the squeezed-limit bispectrum of the cosmic microwave
  background}},  {\em JCAP} {\bf 1305} (2013) 002,
  [\href{http://xxx.lanl.gov/abs/1302.3056}{{\tt arXiv:1302.3056}}].

\bibitem{Kunze:2013hy}
K.~E. Kunze, {\it {CMB and matter power spectra from cross correlations of
  primordial curvature and magnetic fields}},  {\em Phys. Rev.} {\bf D87} (May,
  2013) 103005, [\href{http://xxx.lanl.gov/abs/1301.6105}{{\tt
  arXiv:1301.6105}}].

\bibitem{Subramanian:1998fn}
K.~Subramanian and J.~D. Barrow, {\it {Microwave background signals from
  tangled magnetic fields}},  {\em Phys.Rev.Lett.} {\bf 81} (1998) 3575--3578,
  [\href{http://xxx.lanl.gov/abs/astro-ph/9803261}{{\tt astro-ph/9803261}}].

\bibitem{Mack:2001gc}
A.~Mack, T.~Kahniashvili, and A.~Kosowsky, {\it {Vector and Tensor Microwave
  Background Signatures of a Primordial Stochastic Magnetic Field}},  {\em
  Phys. Rev.} {\bf D65} (2002) 123004,
  [\href{http://xxx.lanl.gov/abs/astro-ph/0105504}{{\tt astro-ph/0105504}}].

\bibitem{Subramanian:2002nh}
K.~Subramanian and J.~D. Barrow, {\it {Small-scale microwave background
  anisotropies due to tangled primordial magnetic fields}},  {\em
  Mon.Not.Roy.Astron.Soc.} {\bf 335} (2002) L57,
  [\href{http://xxx.lanl.gov/abs/astro-ph/0205312}{{\tt astro-ph/0205312}}].

\bibitem{Subramanian:2003sh}
K.~Subramanian, T.~Seshadri, and J.~Barrow, {\it {Small - scale CMB
  polarization anisotropies due to tangled primordial magnetic fields}},  {\em
  Mon.Not.Roy.Astron.Soc.} {\bf 344} (2003) L31,
  [\href{http://xxx.lanl.gov/abs/astro-ph/0303014}{{\tt astro-ph/0303014}}].

\bibitem{Giovannini:2005jw}
M.~Giovannini, {\it {Magnetized cmb anisotropies}},  {\em Class.Quant.Grav.}
  {\bf 23} (2006) R1, [\href{http://xxx.lanl.gov/abs/astro-ph/0508544}{{\tt
  astro-ph/0508544}}].

\bibitem{Paoletti:2008ck}
D.~Paoletti, F.~Finelli, and F.~Paci, {\it {The full contribution of a
  stochastic background of magnetic fields to CMB anisotropies}},  {\em Mon.
  Not. Roy. Astron. Soc.} {\bf 396} (2009) 523--534,
  [\href{http://xxx.lanl.gov/abs/0811.0230}{{\tt arXiv:0811.0230}}].

\bibitem{Yamazaki:2008gr}
D.~G. Yamazaki, K.~Ichiki, T.~Kajino, and G.~J. Mathews, {\it {Effects of a
  Primordial Magnetic Field on Low and High Multipoles of the CMB}},  {\em
  Phys.Rev.} {\bf D77} (2008) 043005,
  [\href{http://xxx.lanl.gov/abs/0801.2572}{{\tt arXiv:0801.2572}}].

\bibitem{Shaw:2009nf}
J.~R. Shaw and A.~Lewis, {\it {Massive Neutrinos and Magnetic Fields in the
  Early Universe}},  {\em Phys. Rev.} {\bf D81} (2010) 043517,
  [\href{http://xxx.lanl.gov/abs/0911.2714}{{\tt arXiv:0911.2714}}].

\bibitem{Giovannini:2009ts}
M.~Giovannini, {\it {Parameter dependence of magnetized CMB observables}},
  {\em Phys.Rev.} {\bf D79} (2009) 103007,
  [\href{http://xxx.lanl.gov/abs/0903.5164}{{\tt arXiv:0903.5164}}].

\bibitem{Paoletti:2010rx}
D.~Paoletti and F.~Finelli, {\it {CMB Constraints on a Stochastic Background of
  Primordial Magnetic Fields}},  {\em Phys.Rev.} {\bf D83} (2011) 123533,
  [\href{http://xxx.lanl.gov/abs/1005.0148}{{\tt arXiv:1005.0148}}].

\bibitem{Shaw:2010ea}
J.~R. Shaw and A.~Lewis, {\it {Constraining Primordial Magnetism}},  {\em
  Phys.Rev.} {\bf D86} (2012) 043510,
  [\href{http://xxx.lanl.gov/abs/1006.4242}{{\tt arXiv:1006.4242}}].

\bibitem{Paoletti:2012bb}
D.~Paoletti and F.~Finelli, {\it {Constraints on a Stochastic Background of
  Primordial Magnetic Fields with WMAP and South Pole Telescope data}},
  \href{http://xxx.lanl.gov/abs/1208.2625}{{\tt arXiv:1208.2625}}.

\bibitem{Ade:2013lta}
{\bf Planck Collaboration} Collaboration, P.~Ade {\em et.~al.}, {\it {Planck
  2013 results. XVI. Cosmological parameters}},
  \href{http://xxx.lanl.gov/abs/1303.5076}{{\tt arXiv:1303.5076}}.

\bibitem{Brown:2005kr}
I.~Brown and R.~Crittenden, {\it {Non-Gaussianity from Cosmic Magnetic
  Fields}},  {\em Phys. Rev.} {\bf D72} (2005) 063002,
  [\href{http://xxx.lanl.gov/abs/astro-ph/0506570}{{\tt astro-ph/0506570}}].

\bibitem{Seshadri:2009sy}
T.~R. Seshadri and K.~Subramanian, {\it {CMB bispectrum from primordial
  magnetic fields on large angular scales}},  {\em Phys. Rev. Lett.} {\bf 103}
  (2009) 081303, [\href{http://xxx.lanl.gov/abs/0902.4066}{{\tt
  arXiv:0902.4066}}].

\bibitem{Caprini:2009vk}
C.~Caprini, F.~Finelli, D.~Paoletti, and A.~Riotto, {\it {The cosmic microwave
  background temperature bispectrum from scalar perturbations induced by
  primordial magnetic fields}},  {\em JCAP} {\bf 0906} (2009) 021,
  [\href{http://xxx.lanl.gov/abs/0903.1420}{{\tt arXiv:0903.1420}}].

\bibitem{Cai:2010uw}
R.-G. Cai, B.~Hu, and H.-B. Zhang, {\it {Acoustic signatures in the Cosmic
  Microwave Background bispectrum from primordial magnetic fields}},  {\em
  JCAP} {\bf 1008} (2010) 025, [\href{http://xxx.lanl.gov/abs/1006.2985}{{\tt
  arXiv:1006.2985}}].

\bibitem{Shiraishi:2010yk}
M.~Shiraishi, D.~Nitta, S.~Yokoyama, K.~Ichiki, and K.~Takahashi, {\it {Cosmic
  microwave background bispectrum of vector modes induced from primordial
  magnetic fields}},  {\em Phys. Rev.} {\bf D82} (2010) 121302,
  [\href{http://xxx.lanl.gov/abs/1009.3632}{{\tt arXiv:1009.3632}}].

\bibitem{Kahniashvili:2010us}
T.~Kahniashvili and G.~Lavrelashvili, {\it {CMB two- and three-point
  correlation functions from Alfv\'en waves}},
  \href{http://xxx.lanl.gov/abs/1010.4543}{{\tt arXiv:1010.4543}}.

\bibitem{Shiraishi:2011dh}
M.~Shiraishi, D.~Nitta, S.~Yokoyama, K.~Ichiki, and K.~Takahashi, {\it {Cosmic
  microwave background bispectrum of tensor passive modes induced from
  primordial magnetic fields}},  {\em Phys. Rev.} {\bf D83} (2011) 123003,
  [\href{http://xxx.lanl.gov/abs/1103.4103}{{\tt arXiv:1103.4103}}].

\bibitem{Shiraishi:2011fi}
M.~Shiraishi, D.~Nitta, S.~Yokoyama, K.~Ichiki, and K.~Takahashi, {\it
  {Computation approach for CMB bispectrum from primordial magnetic fields}},
  {\em Phys. Rev.} {\bf D83} (2011) 123523,
  [\href{http://xxx.lanl.gov/abs/1101.5287}{{\tt arXiv:1101.5287}}].

\bibitem{Shiraishi:2012rm}
M.~Shiraishi, D.~Nitta, S.~Yokoyama, and K.~Ichiki, {\it {Optimal limits on
  primordial magnetic fields from CMB temperature bispectrum of passive
  modes}},  {\em JCAP} {\bf 1203} (2012) 041,
  [\href{http://xxx.lanl.gov/abs/1201.0376}{{\tt arXiv:1201.0376}}].

\bibitem{Trivedi:2010gi}
P.~Trivedi, K.~Subramanian, and T.~R. Seshadri, {\it {Primordial Magnetic Field
  Limits from Cosmic Microwave Background Bispectrum of Magnetic Passive Scalar
  Modes}},  {\em Phys. Rev.} {\bf D82} (2010) 123006,
  [\href{http://xxx.lanl.gov/abs/1009.2724}{{\tt arXiv:1009.2724}}].

\bibitem{Shiraishi:2013wua}
M.~Shiraishi and T.~Sekiguchi, {\it {First observational constraints on tensor
  non-Gaussianity sourced by primordial magnetic fields from cosmic microwave
  background}},  \href{http://xxx.lanl.gov/abs/1304.7277}{{\tt
  arXiv:1304.7277}}.

\bibitem{Babich:2004yc}
D.~Babich and M.~Zaldarriaga, {\it {Primordial bispectrum information from CMB
  polarization}},  {\em Phys.Rev.} {\bf D70} (2004) 083005,
  [\href{http://xxx.lanl.gov/abs/astro-ph/0408455}{{\tt astro-ph/0408455}}].

\bibitem{Yadav:2007rk}
A.~P. Yadav, E.~Komatsu, and B.~D. Wandelt, {\it {Fast Estimator of Primordial
  Non-Gaussianity from Temperature and Polarization Anisotropies in the Cosmic
  Microwave Background}},  {\em Astrophys.J.} {\bf 664} (2007) 680--686,
  [\href{http://xxx.lanl.gov/abs/astro-ph/0701921}{{\tt astro-ph/0701921}}].

\bibitem{Yadav:2007ny}
A.~P. Yadav, E.~Komatsu, B.~D. Wandelt, M.~Liguori, F.~K. Hansen, {\em
  et.~al.}, {\it {Fast Estimator of Primordial Non-Gaussianity from Temperature
  and Polarization Anisotropies in the Cosmic Microwave Background II: Partial
  Sky Coverage and Inhomogeneous Noise}},  {\em Astrophys.J.} {\bf 678} (2008)
  578--582, [\href{http://xxx.lanl.gov/abs/0711.4933}{{\tt arXiv:0711.4933}}].

\bibitem{:2006uk}
{\bf Planck} Collaboration, {\it {Planck: The scientific programme}},
  \href{http://xxx.lanl.gov/abs/astro-ph/0604069}{{\tt astro-ph/0604069}}.

\bibitem{Andre:2013afa}
{\bf PRISM Collaboration} Collaboration, P.~Andre {\em et.~al.}, {\it {PRISM
  (Polarized Radiation Imaging and Spectroscopy Mission): A White Paper on the
  Ultimate Polarimetric Spectro-Imaging of the Microwave and Far-Infrared
  Sky}},  \href{http://xxx.lanl.gov/abs/1306.2259}{{\tt arXiv:1306.2259}}.

\bibitem{Pritchard:2004qp}
J.~R. Pritchard and M.~Kamionkowski, {\it {Cosmic microwave background
  fluctuations from gravitational waves: An analytic approach}},  {\em Annals
  Phys.} {\bf 318} (2005) 2--36,
  [\href{http://xxx.lanl.gov/abs/astro-ph/0412581}{{\tt astro-ph/0412581}}].

\bibitem{Kamionkowski:2010rb}
M.~Kamionkowski and T.~Souradeep, {\it {The Odd-Parity CMB Bispectrum}},  {\em
  Phys. Rev.} {\bf D83} (2011) 027301,
  [\href{http://xxx.lanl.gov/abs/1010.4304}{{\tt arXiv:1010.4304}}].

\bibitem{Shiraishi:2011st}
M.~Shiraishi, D.~Nitta, and S.~Yokoyama, {\it {Parity Violation of Gravitons in
  the CMB Bispectrum}},  {\em Prog.Theor.Phys.} {\bf 126} (2011) 937--959,
  [\href{http://xxx.lanl.gov/abs/1108.0175}{{\tt arXiv:1108.0175}}].

\bibitem{Shiraishi:2012sn}
M.~Shiraishi, {\it {Parity violation of primordial magnetic fields in the CMB
  bispectrum}},  {\em JCAP} {\bf 1206} (2012) 015,
  [\href{http://xxx.lanl.gov/abs/1202.2847}{{\tt arXiv:1202.2847}}].

\bibitem{Smith:2006ud}
K.~M. Smith and M.~Zaldarriaga, {\it {Algorithms for bispectra: Forecasting,
  optimal analysis, and simulation}},  {\em Mon.Not.Roy.Astron.Soc.} {\bf 417}
  (2011) 2--19, [\href{http://xxx.lanl.gov/abs/astro-ph/0612571}{{\tt
  astro-ph/0612571}}].

\bibitem{Hanson:2009kg}
D.~Hanson, K.~M. Smith, A.~Challinor, and M.~Liguori, {\it {CMB lensing and
  primordial non-Gaussianity}},  {\em Phys.Rev.} {\bf D80} (2009) 083004,
  [\href{http://xxx.lanl.gov/abs/0905.4732}{{\tt arXiv:0905.4732}}].

\bibitem{Mangilli:2009dr}
A.~Mangilli and L.~Verde, {\it {Non-Gaussianity and the CMB Bispectrum:
  confusion between Primordial and Lensing-Rees Sciama contribution?}},  {\em
  Phys.Rev.} {\bf D80} (2009) 123007,
  [\href{http://xxx.lanl.gov/abs/0906.2317}{{\tt arXiv:0906.2317}}].

\bibitem{Lewis:2011fk}
A.~Lewis, A.~Challinor, and D.~Hanson, {\it {The shape of the CMB lensing
  bispectrum}},  {\em JCAP} {\bf 1103} (2011) 018,
  [\href{http://xxx.lanl.gov/abs/1101.2234}{{\tt arXiv:1101.2234}}].

\bibitem{Pearson:2012ba}
R.~Pearson, A.~Lewis, and D.~Regan, {\it {CMB lensing and primordial squeezed
  non-Gaussianity}},  {\em JCAP} {\bf 1203} (2012) 011,
  [\href{http://xxx.lanl.gov/abs/1201.1010}{{\tt arXiv:1201.1010}}].

\bibitem{Junk:2012qt}
V.~Junk and E.~Komatsu, {\it {Cosmic Microwave Background Bispectrum from the
  Lensing--Rees-Sciama Correlation Reexamined: Effects of Non-linear Matter
  Clustering}},  {\em Phys.Rev.} {\bf D85} (2012) 123524,
  [\href{http://xxx.lanl.gov/abs/1204.3789}{{\tt arXiv:1204.3789}}].

\bibitem{Lewis:2012tc}
A.~Lewis, {\it {The full squeezed CMB bispectrum from inflation}},  {\em JCAP}
  {\bf 1206} (2012) 023, [\href{http://xxx.lanl.gov/abs/1204.5018}{{\tt
  arXiv:1204.5018}}].

\bibitem{Hu:2012td}
B.~Hu, M.~Liguori, N.~Bartolo, and S.~Matarrese, {\it {Future CMB ISW-Lensing
  bispectrum constraints on modified gravity in the Parameterized
  Post-Friedmann formalism}},  {\em Phys.Rev.} {\bf D88} (2013) 024012,
  [\href{http://xxx.lanl.gov/abs/1211.5032}{{\tt arXiv:1211.5032}}].

\bibitem{DiValentino:2012yg}
E.~Di~Valentino, A.~Melchiorri, V.~Salvatelli, and A.~Silvestri, {\it
  {Parametrised modified gravity and the CMB Bispectrum}},  {\em Phys.Rev.}
  {\bf D86} (2012) 063517, [\href{http://xxx.lanl.gov/abs/1204.5352}{{\tt
  arXiv:1204.5352}}].

\bibitem{Hanson:2013hsb}
{\bf SPTpol Collaboration} Collaboration, D.~Hanson {\em et.~al.}, {\it
  {Detection of B-mode Polarization in the Cosmic Microwave Background with
  Data from the South Pole Telescope}},
  \href{http://xxx.lanl.gov/abs/1307.5830}{{\tt arXiv:1307.5830}}.

\bibitem{Hu:2001fb}
W.~Hu, {\it {Dark synergy: Gravitational lensing and the CMB}},  {\em
  Phys.Rev.} {\bf D65} (2002) 023003,
  [\href{http://xxx.lanl.gov/abs/astro-ph/0108090}{{\tt astro-ph/0108090}}].

\bibitem{Santos:2013gqa}
L.~Santos, P.~Cabella, A.~Balbi, and N.~Vittorio, {\it {Neutrinos and dark
  energy constraints from future galaxy surveys and CMB lensing information}},
  \href{http://xxx.lanl.gov/abs/1307.2919}{{\tt arXiv:1307.2919}}.

\end{thebibliography}\endgroup
\end{document}